\newif\ifpreprint
\preprinttrue

\documentclass{osa-article}

\usepackage{subcaption}
\usepackage{float}

\ifpreprint
\journal{preprint_mb}
\else
\journal{oe}
\fi

\begin{document}

\title{Inverse Design of Nanophotonic Devices using Dynamic Binarization}

\author{Marco Butz,\authormark{1,2,3} Adrian S. Abazi,\authormark{1,2,3} Rene Ross,\authormark{1,2,3} Benjamin Risse,\authormark{4,5} and Carsten Schuck\authormark{1,2,3, *}}

\address{
\authormark{1}Institute of Physics, University of Münster, Wilhelm-Klemm-Str. 10, 48149 Münster, Germany\\
\authormark{2}Center for NanoTechnology (CeNTech), Heisenbergstr. 11, 48149 Münster, Germany\\
\authormark{3}Center for Soft Nanoscience (SoN), Busso-Peus-Str. 10, 48149 Münster, Germany\\
\authormark{4}Institute for Geoinformatics, University of Münster, Heisenbergstr. 2, 48149 Münster, Germany\\
\authormark{5}Institute for Computer Science, University of Münster, Einsteinstraße 62, 48149 Münster, Germany}

\email{\authormark{*}carsten.schuck@uni-muenster.de}



\begin{abstract}
The complexity of applications addressed with photonic integrated circuits is steadily rising and poses increasingly challenging demands on individual component functionality, performance and footprint. Inverse design methods have recently shown great promise to address these demands using fully automated design procedures that enable access to non-intuitive device layouts beyond conventional nanophotonic design concepts. Here we present a dynamic binarization method for the objective-first algorithm that lies at the core of the currently most successful inverse design algorithms. Our results demonstrate significant performance advantages over previous implementations of objective first algorithms, which we show for a fundamental TE$_{00}$ to TE$_{20}$ waveguide mode converter both in simulation and in experiments with fabricated devices. 
\end{abstract}
%above is version as suggested by benjamin (kind of)
%old version:
%The complexity of applications addressed with photonic integrated circuits is steadily rising and poses increasingly challenging demands on individual component functionality, performance and footprint. Inverse design methods have recently shown great promise to address these demands using fully automated design procedures that enable access to non-intuitive device layouts beyond conventional nanophotonic design concepts. Here we demonstrate a significant improvement of the objective-first method that lies at the core of the currently most successful inverse design algorithms. We reach improved device performance by removing structural biases and implementing a dynamic binarization method during optimization. The performance advantages over previous implementations of objective first algorithms are shown for a fundamental TE$_{00}$ to TE$_{20}$ waveguide mode converter both in simulation and in experiments with fabricated devices.
%%%%%%%%%%%%%%%%%%%%%%%%%%  body  %%%%%%%%%%%%%%%%%%%%%%%%%%
\section{Introduction}
The design process of photonic integrated circuit components typically relies on brute force optimization of critical parameters in an initial layout that was based on some physical intuition deriving from established concepts such as interference, diffraction, coupled mode theory or similar. While a wide range of devices like directional couplers \cite{Trinh1995}, resonators \cite{Bogaerts2012}, photonic crystals\cite{olthaus2020}, adiabatic tapers \cite{Fijol2003} or coupling structures from optical fibers to on-chip integrated photonic waveguides have been created following this paradigm, they all suffer from large device footprints and a limited number of tunable design parameters. The few degrees of freedom available to the designer (typically less than ten) usually only provide access to a tiny fraction of the solution space, which may not contain layouts combining optimal efficiency and small footprint. Nanophotonic circuits offer solutions to problems of ever-increasing complexity \cite{Steinbrecher2019}. In order to guarantee future scalability, fully automated design processes for more efficient and compact devices that exhibit novel functionalities based on complex interference phenomena not accessible by human intuition are needed. \\
Recently, various techniques to design nanophotonic devices following a top-down approach, such as topology optimization \cite{Jensen2005,Frellsen2016,CalaLesina2022}, genetic algorithms \cite{Sanchis2004,Jiang2003}, particle swarm optimization \cite{Mak2016} or the objective-first approach \cite{Lu2012}, have emerged. The objective-first algorithm has proven particularly successful for designing high-efficiency and low-footprint nanophotonic devices \cite{Lu2012,Lu2013,Tutgun2018,Alpklc2018} and currently constitutes one of the most promising nanophotonic inverse design methods.\\
A problem of the objective-first approach however is that the resulting permittivity distribution consists of continuously varying values, while modern nanophotonic fabrication methods only allow for permittivity distributions containing two (few) discrete values, as each position in the design can only be occupied by either one material, e.g. the dielectric, or another, e.g. air or silicon dioxide. In the originally proposed version of the objective first algorithm, reconciliation with a discrete permittivity distribution is accomplished by thresholding the continuous permittivity map. This constitutes a dramatic intervention in the optimization process because thresholding is not being guided by any gradient information. Even subsequently applied adjoint sensitivity analysis \cite{Jensen2005,Lu2013} is inadequate to compensate for the loss of the objective-first-specific “global” optimization characteristic and will only yield solutions that are far from any optimum the objective-first algorithm had initially converged to.\\
Static binarization penalty terms to the objective-first method have been proposed to overcome this problem by steadily increasing the binarity (dielectric vs. air) of the structure with each iteration \cite{Callewaert2016}. While this modification has shown improvements, it tends to limit the ability of the algorithm to develop freely during optimization. The corresponding penalty term uses a thresholded version of the device structure in a previous iteration as a reference for calculating the update of the binarization level, which is why we refer to this technique as the “static” binarization method. We here remove the aforementioned structural bias by calculating the binarization level update in real time in each iteration of the convex optimization step. To do so, we exploit so called mixed-integer optimization capabilities of state-of-the-art convex optimization software. Our “dynamic” binarization method thus removes critical limitations of "static" techniques on the iterative evolution of the device structure towards an efficient solution.\\
We here demonstrate the superior performance of our approach for the example of a spatial waveguide mode converter, transforming fundamental transverse electric modes, TE$_{00}$, into TE$_{20}$ modes, which is challenging to achieve with other design methods. Such mode converters are fundamental components as they are essential for various tasks such as optical fiber coupling \cite{Frish2014a, Fang2010}, phase matching \cite{Penner1994} or coupling between different waveguide geometries \cite{Lu2012}. However, our approach straightforwardly applies to a wide range of other nanophotonic circuit components.\\
We obtain a highly efficient and compact device optimized in full 3D exhibiting a conversion efficiency of 81.1\,\% and a footprint of just 10\,\textmu m$^2$. In direct comparison to a device optimized using the static binarization method we observe an increase of $9.8$\,\% in conversion efficiency. We fabricate the device on a 200\,nm thin Si$_3$N$_4$-on-insulator platform using electron beam lithography and verify its functionality using mode imaging techniques. 

\section{Inverse design algorithm}
\begin{figure}[h]
\centering\includegraphics[width=\textwidth]{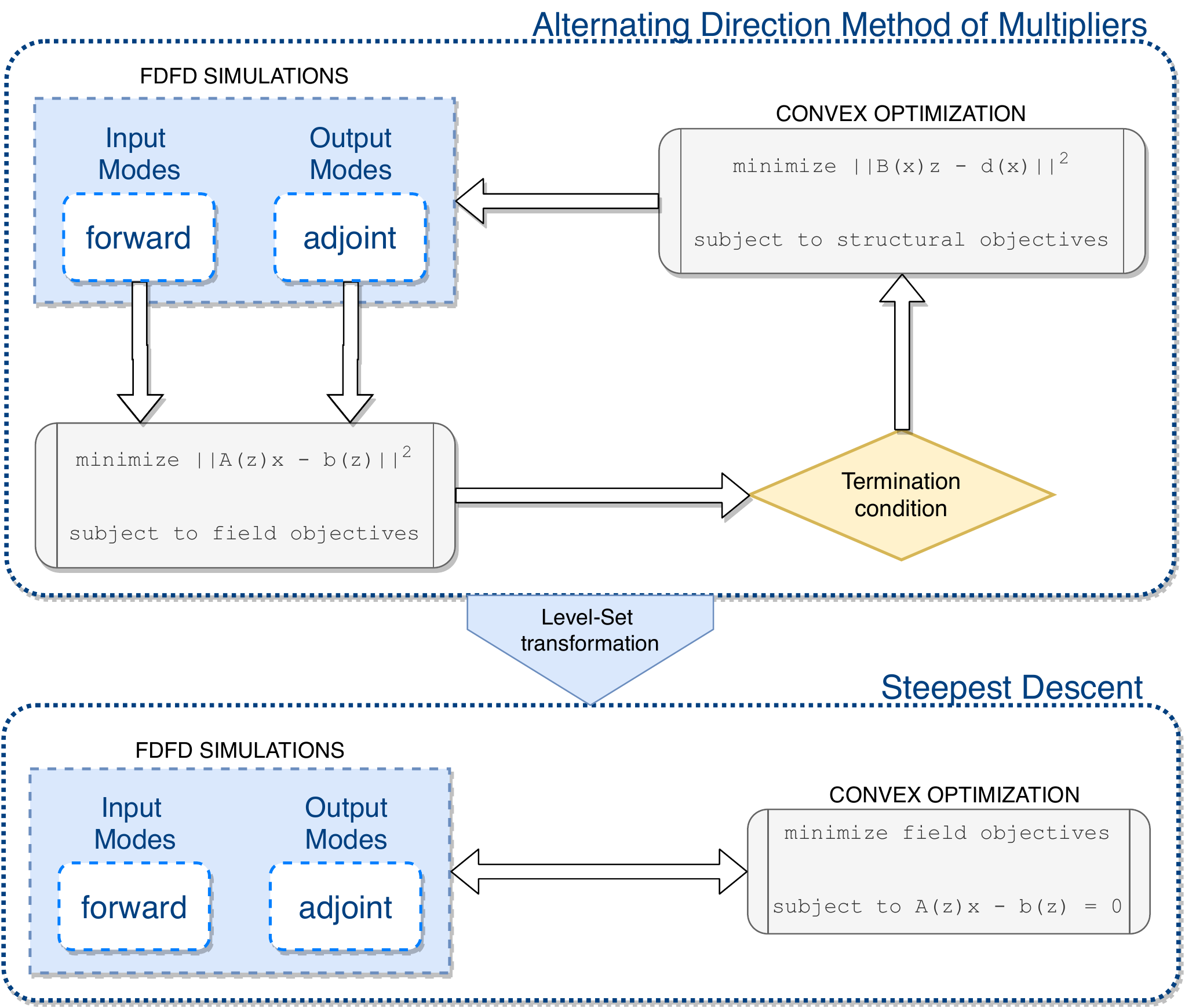}
\caption{Our optimization algorithm can be split into two separate stages. In the first step we optimize the structure with continuous permittivity values using an \textit{Alternating Direction Method of Multipliers}-implementation of the "objective-first" algorithm. We alternate between solving the structural subproblem and solving the field subproblem until a termination criterion is met, which is either tied to the physics residual or the maximum number of iterations. Afterwards, we convert the resulting permittivity distribution into a level-set representation to fine tune the structure using the adjoint sensitivity analysis and following the gradient's steepest descent. The second stage is terminated after a fixed number of iterations.}
\label{fig:flowchart}
\end{figure}
While manual design methods usually follow a bottom-up approach relying predominantly on phenomena accessible by human intuition and brute force parameter optimization, inverse design algorithms follow a top-down approach by specifying given input and desired output waveforms. The design region is considered as a black box, spatially discretized into pixels. The algorithm then tries to find a permittivity distribution, $\epsilon$, within the design region that fulfills the design objectives. The structure of our algorithm is shown in \autoref{fig:flowchart}. We split the optimization problem into two sub-problems, namely the composition of the target field (field sub-problem) and solving for the corresponding structure (structural sub-problem), which enables us to apply the \textit{alternating direction method of multipliers} (ADMM) \cite{Wen2010}. Following the "objective-first" approach, we only allow for small changes of the target field within this process. We derive the two sub-problems starting from a transformation of the time-harmonic Maxwell equations into a linear algebra form, such that 
\begin{align}
    (\nabla \times \mu_0^{-1} \nabla \times - \omega^2 \epsilon)E_i + i \omega J = 0 \\
    \rightarrow A(z)x - b(z) = 0 \label{eq:linearAlgebraMaxwell}
    %show substitutions?
\end{align}
where $\mu_0$ is the permeability, $z$ is representative of the permittivity $\epsilon$, $x$ the electric field $E$, and $b$ depends on the electric current density $J$. The field sub-problem can be expressed as
\begin{equation}
    \centering
    \begin{tabular}{cl}
        $\mathrm{\underset{x}{minimize}}$ & $|| A(z)x - b(z) ||^2$ \\
        $\mathrm{subject\:to}$ & $f(x) = 0$
    \end{tabular}
    \label{eq:FieldProblem}
\end{equation}
where $f(x)$ is the field objective function. In each iteration, the starting point for the convex optimization in \autoref{eq:FieldProblem} is determined by combining finite-difference frequency-domain (FDFD) simulations driven by the specified input-modes and adjoint-FDFD simulations driven by the specified output-waveforms. The structural sub-problem, where we aim to obtain a corresponding permittivity distribution, can be expressed in a similar way: 
\begin{equation}
    \centering
    \begin{tabular}{cl}
        $\mathrm{\underset{z}{minimize}}$ & $|| B(x)z - d(x) ||^2$ \\
        $\mathrm{subject\:to}$ & $z_{\mathrm{min}} < z < z_{\mathrm{max}}$
    \end{tabular}
    \label{eq:StructureProblem}
\end{equation}
To obtain \autoref{eq:StructureProblem} we utilize the bi-linearity of \autoref{eq:linearAlgebraMaxwell}, and find that $A(z)x - b(z) = B(x)z - d(x)$. Here, the structure parameter $z$ is constrained to reside between $z_{\mathrm{min}}=0$ and $z_{\mathrm{max}}=1$, which are linearly mapped to the material-dependent minimum and maximum permittivity $\epsilon_{\mathrm{min}}$ and $\epsilon_{\mathrm{max}}$, respectively.
%It is always ensured that the specified design objectives $f(x)$ (i.e. a specific waveform at the output port of the device - see \autoref{sec:problemSpecificSection} for details) are fulfilled throughout the optimization procedure. 
Enforcing the specified design objectives $f(x)$ in \autoref{eq:FieldProblem} leads to \autoref{eq:linearAlgebraMaxwell} not being fulfilled. The residual of this function ("physics residual") is the figure of merit of the optimization procedure. This method can be extended to support multiple objectives by transforming \autoref{eq:linearAlgebraMaxwell} into a sum over all desired objectives involved.  
%show final minimization problem? -- CS: ADMM sollte man hier vor allem deswegen erwähnen weil es in Fig. 1 sehr explizit auftaucht. Bitte füge zu ADMM noch 1-2 Sätze ein aber ich würde hier nicht mit Lagrangians etc. anfangen. Der Text sollte enger auf Fig. 1 eingehen - z.B. sprichst Du in der Caption von field und structural sub-problems: diese Sprache musst Du dann auch im Main Text übernehmen. 
%MB: Gelöst mit commits am 17.11.22?
\\
After this initial optimization step, we convert the structure to a boundary parameterization using a level-set representation and fine tune the device following a first order gradient in the direction of steepest descent. \cite{Giles2000}
\section{Binarization of the continuous structure}
The objective-first optimization strategy suffers from a significant drawback. In its original form \cite{Lu2012} the resulting structure features continuous permittivity values ranging from user-specified minimum to maximum permittivities. As typically only distributions containing two discrete permittivity values can be realized in most modern nanofabrication processes, such as electron beam lithography, the resulting structure has to be converted from continuous to binary values.\\
The initial solution \cite{Lu2012} was to threshold the continuous distribution, i.e. to introduce a critical permittivity value between $\epsilon_{\mathrm{min}}$ and $\epsilon_{\mathrm{max}}$. Each $\epsilon$ that lies below the critical value is being set to $\epsilon_{\mathrm{min}}$ and each value that lies above it is being set to $\epsilon_{\mathrm{max}}$. As this is a critical intervention in the optimization flow, which is not based on any gradient information, the resulting structure can be expected to differ significantly from the result of the first optimization step in both permittivity distribution and performance. \\
Efforts to overcome this issue have relied on introducing a static binarization penalty term to \autoref{eq:linearAlgebraMaxwell}, such that the minimization task can be expressed as \cite{Callewaert2016}
\begin{equation}
    \centering
    \begin{tabular}{cl}
        $\mathrm{\underset{z}{minimize}}$ & $|| B(x)z - d(x) + \lambda z - z_{\mathrm{bin}} ||$ \\
        $\mathrm{subject\:to}$ & $z_{\mathrm{min}} < z < z_{\mathrm{max}}$
    \end{tabular}
    \label{eq:staticBinarization}
\end{equation}
where $\lambda$ is the binarization strength, which steadily increases throughout the optimization and $z_\mathrm{bin}$ is the binarized version of the distribution obtained in the previous iteration via thresholding. While this approach reduces the impact of the critical thresholding step on the final structure, it significantly limits the structure's ability to develop freely. By obtaining $z_{\mathrm{bin}}$ based on the previous structure and keeping it constant while solving the minimization problem (\ref{eq:staticBinarization}), each pixel tends to develop into a given direction determined by $z_\mathrm{bin}$. The result is a static behavior, where drastic but beneficial structural changes that might occur especially in later stages of the optimization process will thus be suppressed. \\
To remove this structural bias, we transform $z_\mathrm{bin}$ into an optimization variable that is constrained to binary values and make it accessible to the convex optimization toolkit, which is used for solving \autoref{eq:staticBinarization}. This enables us to also formulate the binarization as a hard constraint enforcing a certain level of binarity in each iteration:
\begin{equation}
    \centering
    \begin{tabular}{cl}
        $\mathrm{\underset{z}{minimize}}$ & $|| B(x)z - d(x) ||$ \\
        $\mathrm{subject\:to}$ & $z_{\mathrm{min}} < z < z_{\mathrm{max}}$\\
        & $\lambda \cdot N \geq || z - z_\mathrm{bin} ||$
    \end{tabular}
    \label{eq:dynamicBinarizationHard}
\end{equation}
Here $\lambda$ is being decreased during the optimization process with $\lambda = 1$ allowing an unrestricted continuous structure and $\lambda=0$ enforcing completely binary structure variables. To technically implement the hard constraint we introduced the normalization term $N=||z_\mathrm{even}||$ with $z_\mathrm{even}$ representing evenly distributed structure variables $z_i=0.5$.
\section{Simulation procedure and computational methods}
To calculate the electromagnetic target field within the device that we aim to find a correspondent permittivity distribution for, we need to perform two electromagnetic simulations for each input mode involved. One forward simulation using the specified input mode as a source and an adjoint simulation where the desired output mode provides the source. We simulate these fields using the finite-difference frequency-domain (FDFD) framework \textit{OpenCLFDFD} \cite{Petykiewicz}, which enables hardware accelerated parallel computation. We run the FDFD computations on High Performance Computing resources (Nvidia Tesla A100 GPUs on the PALMA II cluster) to reduce the computation time for each simulation. \\
To solve the minimization problems stated in equations \ref{eq:staticBinarization} and \ref{eq:dynamicBinarizationHard}, we use the Matlab package \textit{CVX} \cite{cvx}. To perform convex optimization involving binary variables we employ the mixed-integer optimization capabilities of the \textit{MOSEK} solver \cite{mosek} supported by \textit{CVX}. We configure the solver to use sophisticated heuristics \cite{Bonami2009} to ensure that a good initial solution to the mixed integer problem can be found quickly.

\section{Design of a TE$_{00}$ to TE$_{20}$ waveguide mode converter}
\label{sec:problemSpecificSection}
\begin{figure}[h!]
\centering\includegraphics[width=\textwidth]{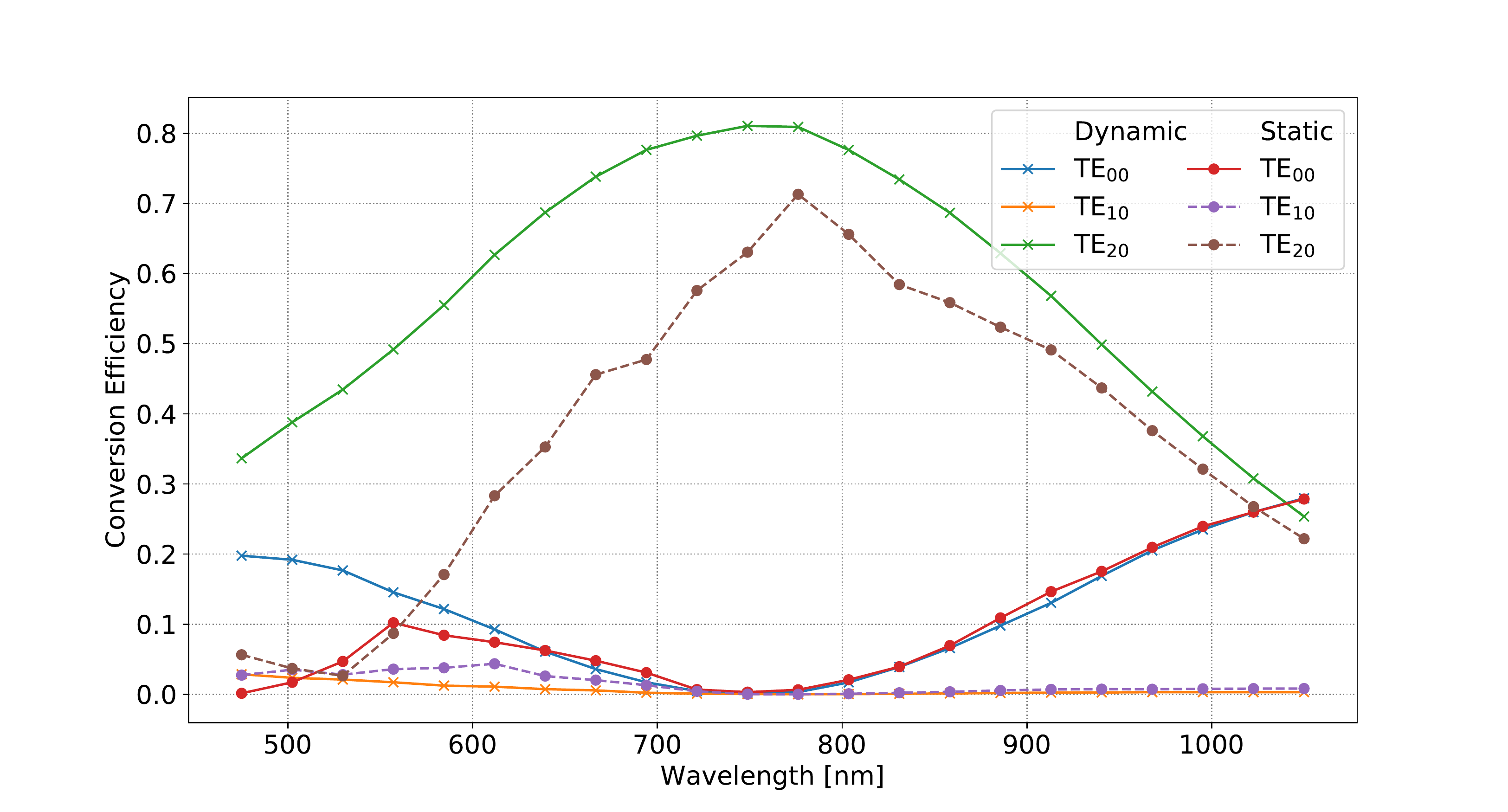}
\caption{Conversion efficiencies of the fundamental TE$_{00}$ input-to lowest-order output modes, simulated for a range of wavelengths centered around the design wavelength of $775$\,nm. Results for the device optimized using the dynamic and static binarization methods are depicted with the solid and dashed lines, respectively.}
\label{fig:broadband}
\end{figure}

\begin{figure}[H]
\centering\includegraphics[width=\textwidth]{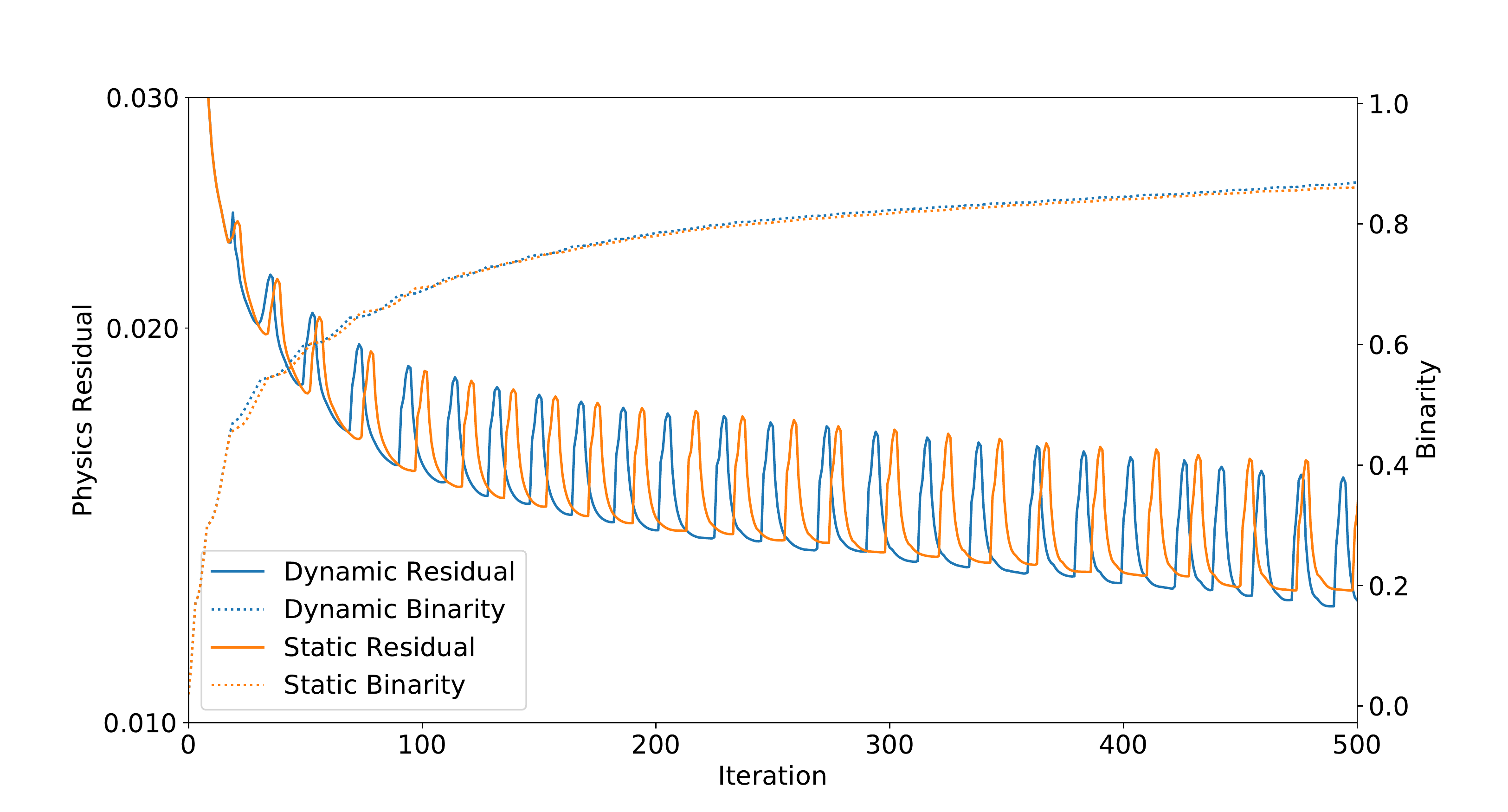}
\caption{Evolution of the physics residual during the optimization for both the dynamic and static binarization methods, implemented as a soft constraint with the corresponding binarity of the structure. The oscillations in the residuals are related to structure resets because the algorithm did not succeed in lowering the residual in a certain direction of the optimization.}
\label{fig:residuals}
\end{figure}

To demonstrate the benefits of using dynamic over static binarization functions, we design a TE$_{00}$ to TE$_{20}$ waveguide mode converter for a wavelength of $775$\,nm in full 3D. We chose 200\,nm thin silicon nitride (Si$_3$N$_4$) on 2\,\textmu m thick silicon dioxide (SiO$_2$) as a material platform. The pixel size is set to 40\,nm and the initial structure is composed of uniformly distributed permittivity values, i.e. $z_i = 0.5$. We specify the device width and length as 3.64\,\textmu m and 2.84\,\textmu m, respectively. We conducted 200 iterations of the objective-first step followed by 200 iterations using the steepest-descent method. The resulting performance of devices simulated simulated with static and dynamic binarization are shown in \autoref{fig:broadband}.
The data show that the conversion efficiency into the TE$_{20}$ mode of the dynamically binarized device is higher over the entire simulated range from $475$\,nm to $1025$\,nm wavelength. While the device was only optimized for the design wavelength of $775$\,nm, its $3$\,dB bandwidth ranges from $575$\,nm to $925$\,nm. We further observe that the dynamic binarization method is capable of reducing the physics residual more efficiently than static binarization approaches. \autoref{fig:residuals} compares the evolution of the physics residual for a soft constraint implementation (following \autoref{eq:staticBinarization}) of a dynamic and a static binarization term.
We find that, especially in later stages of the optimization, the dynamic binarization term is able to lower the physics residual further than the static version while even producing a slightly higher level of binarity. Note that small differences in the physics residual can have a significant impact in the final device performance because the result of this optimization step is used as an input to fine tune the structure using a steepest descent method.
\section{Fabrication and experimental verification}
\begin{figure}[H]
    \includegraphics[width=\textwidth]{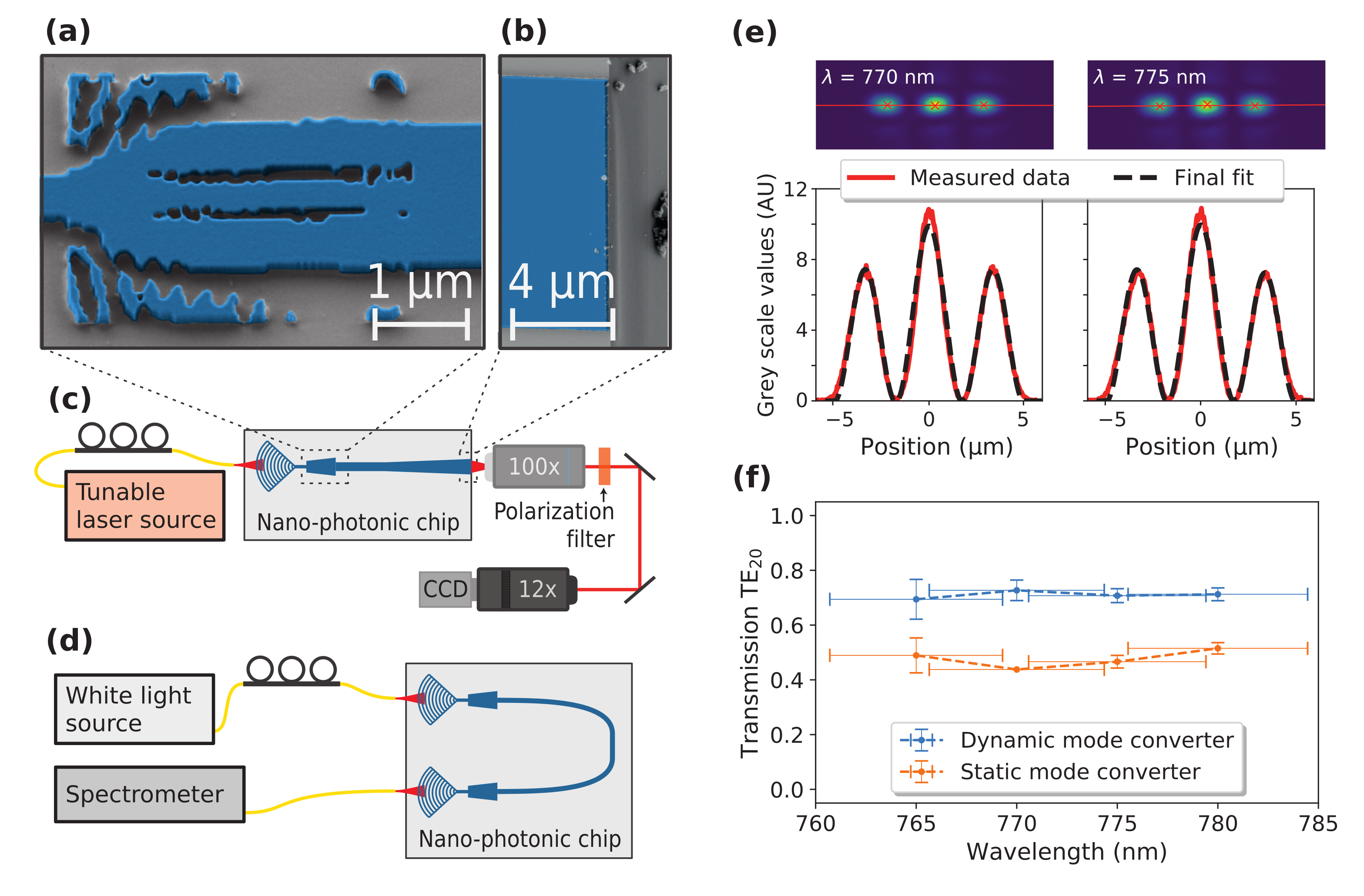}
    \caption{\textbf{(a)} False-color scanning electron micrograph of fabricated SiN-on-insulator mode converters, optimized using a dynamic binarization function. \textbf{(b)} Cleaved edge of the chip with waveguide facet after adiabatic tapering. \textbf{(c)} Measurement setup to determine the relative mode composition using a CCD-camera and magnification lenses. \textbf{(d)} Setup to determine the insertion loss per device. \textbf{(e)} Mode profiles with intensity distribution measured along the red lines. The fit used to numerically determine the mode composition is visualized using a dashed line overlay. \textbf{(f)} Measured absolute transmission of the TE$_{20}$ mode for the dynamically and statically binarized devices.}
    \label{fig:fab_experiment}
\end{figure}

To verify the functionality of the device, we fabricate the calculated binary permittivity distributions on a $200$\,nm silicon nitride on $2$\,\textmu m silicon dioxide on silicon platform using high-resolution electron beam lithography (EBPG 5150, Raith) and reactive ion etching. \autoref{fig:fab_experiment} (a) shows a false-color scanning electron micro-graph of the fabricated device, which has been optimized using the dynamic binarization method. The output waveguide is tapered up adiabatically until it reaches a width of $10$\,\textmu m, thus also enlarging the mode profile while maintaining the mode composition. The enlarged profile enables us to examine the mode using a microscope objective after it has been emitted to free space. Here, the undisturbed emission of the waveguide mode is achieved by cleaving the nanophotonic chip orthogonal to the output waveguide such that we obtain a clean waveguide facet as depicted in \autoref{fig:fab_experiment} (b). As shown in \autoref{fig:fab_experiment} (c), the device is illuminated through a fiber-coupled and polarization-controlled Ti:Sapphire laser at $\lambda=775$\,nm connected to a fiber array by single-mode fibers which is interfaced to the nanophpotonic chip via grating couplers. The filling factor and period of the coupling structures are optimized for TE$_{00}$ modes at $\lambda=775$\,nm. A 100x microscope objective is used to further enlarge the emitted mode profile and collect the light via a CCD-camera after a second 12x magnifying lens configuration. A polarization filter between the aforementioned magnification components allows us to filter out either the horizontal $H$- or vertical $V$-components of the electric field vector. 
% please switch c and d labels in figure and then also here in the text so all labels appear in alphabetical order. 
Using this setup, we observe the filtered intensity distributions corresponding to the superposition of either the $H$- or $V$-components of the modes. \autoref{fig:fab_experiment} (e) shows these intensity distributions, each of them featuring three maxima, which are qualitatively consistent with the simulated results.\\
%fit:
We utilize a least square fitting algorithm to match simulated values of intensity profiles to the measured data, indicated by red horizontal lines, see \autoref{fig:fab_experiment} (e). The amplitudes and relative phases of the individually calculated waveguide modes are the free parameters determined by the fitting algorithm, allowing for the quantitative determination of the mode purity of the fabricated device's output.\\
To determine the insertion loss of the device, we employ the measurement setup shown in \autoref{fig:fab_experiment} (d). A polarization-controlled and single-mode fiber-coupled white light source is used to illuminate the input grating coupler through a fiber array, while the output port is analyzed by a spectrometer. The circuitry consists of two identical mode conversion devices which are connected through their output-waveguides. The second converter reverses the mode conversion process such that the insertion losses of the two devices accumulate. After measuring the insertion losses of the grating couplers in similar reference circuits, we hereby may determine the total losses of an individual mode converter. 
%frage zu folgendem?
To ensure compatibility between the total transmission of the reference devices and the relative mode composition of the measured device, we determine the mean total transmission and its standard deviation of the mode converter and the grating coupler for multiple fabricated circuits.
Combining the hereby measured mean insertion loss with the relative mode composition determined by measuring the intensity profile recorded by the CCD-camera, we can calculate the absolute conversion efficiency into the TE$_{20}$ mode.\\
Though limited by the narrow effective wavelength range of the employed grating couplers, we find comparable performance differences between the statically and dynamically binarized device layouts. The data is visualized in \autoref{fig:fab_experiment} (f), albeit we observe slightly lower efficiencies than expected based on the simulations depicted in \autoref{fig:broadband}, likely due to fabrication imperfections. The uncertainty in wavelength results from the FWHM of the Ti:Sapphire laser, while the uncertainty of the Transmission into the TE$_{20}$ mode results from the standard deviation of the fitting parameters, as well as the standard deviations of the mean total transmission of identical mode converters and grating couplers. Our results are consistent with significantly enhanced conversion efficiencies from the TE$_{00}$ into the TE$_{20}$ mode for dynamically binarized devices compared to their statically binarized counterparts.
%old figure:
\iffalse
\begin{figure}[h!]
\begin{subfigure}{.5\textwidth}
  \centering
  \includegraphics[width=.6\textwidth]{img/field_sim.png}
  \caption{}
  \label{fig:structure}
\end{subfigure}%
\begin{subfigure}{.5\textwidth}
  \centering
  \includegraphics[width=\textwidth]{img/sem_mode.png}
  \caption{}
  \label{fig:mode}
\end{subfigure}%
\caption{Simulated device with simulated field overlay showing the E_$_y$-component (\textbf{a}). An SEM image of the fabricated device is shown in \textbf{b}. The inlay shows the recorded mode intensity profile at a wavelength of $\lambda=775$\,nm after adiabatically tapering up the output waveguide and collecting the mode shining out of the cleaved chip.}
\end{figure}
\fi
\section{Conclusion}
We have introduced a novel dynamic binarization bias for gradient-based continuous inverse design methods in nanophotonics and demonstrated significant performance improvements over previous static implementations. Our approach leverages state-of-the-art mixed-integer optimization techniques and is integrated into established objective-first inverse design frameworks with only minor modifications, while not imposing any new application constraints. As a specific example of a non-trivial photonic integrated circuit component, we show the optimization of a TE$_{00}$ $\rightarrow$ TE$_{20}$ waveguide mode converter in full 3D. Our dynamic binarization method achieves a $9.8$\,\% increase in conversion efficiency compared to the currently used static binarization approaches. We fabricated and experimentally tested the functionality of the device through direct imaging of the waveguide mode after adiabatic tapering and find qualitative agreement with the simulated conversion efficiency, thus confirming the superiority of the dynamically binarized device. Our work combines progress in two rapidly evolving fields of research, namely mixed-integer optimization and nanophotonic inverse design, thus highlighting the potential for improving photonic integrated circuit performance. We anticipate that our work benefits a wide range of applications in nanophotonics and integrated quantum technology \cite{moody2022}. 

\section*{Funding}
Ministerium für Kultur und Wissenschaft des Landes Nordrhein-Westfalen (421-8.03.03.02–130428).
German Research Foundation (CRC 1459, C05).

\section*{Acknowledgments}
We would like to thank the Münster Nanofabrication Facility (MNF) for their support in nanofabrication related matters. C.S. acknowledges support from the Ministry for Culture and Science of North Rhine-Westphalia (421-8.03.03.02–130428). The authors acknowledge support by the German Research Foundation (DFG, CRC 1459).

\section*{Disclosures}
The authors declare no conflicts of interest.

\bibliography{library}

\end{document}